\documentclass[aps,prl,preprint,tightenlines,superscriptaddress,showpacs,byrevtex]{revtex4}
\usepackage{epsfig}
\usepackage{wrapfig,graphicx} 
\usepackage{dcolumn}  
\renewcommand{\thefootnote}{\fnsymbol{footnote}} 
\graphicspath{{ps}}

\begin{document}

\preprint{\vbox{
  \vskip -6mm               
	         \hbox{BELLE-CONF-0349}
                 \hbox{LP03 Parallel Sessions: XX}
                 \hbox{LP03-ID nnn}
                 \hbox{hep-ex nnnn, 
		 \hbox{V2.2}}
}}

\title {Measurement of the Branching Fraction for  B$\rightarrow \, \psi(2S)K^{*}(892)$ Decays} 
\normalsize
\affiliation{Aomori University, Aomori}
\affiliation{Budker Institute of Nuclear Physics, Novosibirsk}
\affiliation{Chiba University, Chiba}
\affiliation{Chuo University, Tokyo}
\affiliation{University of Cincinnati, Cincinnati OH}
\affiliation{Gyeongsang National University, Chinju}
\affiliation{University of Hawaii, Honolulu HI}
\affiliation{High Energy Accelerator Research Organization (KEK), Tsukuba}
\affiliation{Hiroshima Institute of Technology, Hiroshima}
\affiliation{Institute of High Energy Physics, Chinese Academy of Sciences, Beijing}
\affiliation{Institute of High Energy Physics, Vienna}
\affiliation{Institute for Theoretical and Experimental Physics, Moscow}
\affiliation{J. Stefan Institute, Ljubljana}
\affiliation{Kanagawa University, Yokohama}
\affiliation{Korea University, Seoul}
\affiliation{Kyoto University, Kyoto}
\affiliation{Kyungpook National University, Taegu}
\affiliation{Institut de Physique des Hautes \'Energies, Universit\'e de Lausanne, Lausanne}
\affiliation{University of Maribor, Maribor}
\affiliation{University of Melbourne, Victoria}
\affiliation{Nagoya University, Nagoya}
\affiliation{Nara Women's University, Nara}
\affiliation{National Kaohsiung Normal University, Kaohsiung}
\affiliation{National Lien-Ho Institute of Technology, Miao Li}
\affiliation{National Taiwan University, Taipei}
\affiliation{H. Niewodniczanski Institute of Nuclear Physics, Krakow}
\affiliation{Nihon Dental College, Niigata}
\affiliation{Niigata University, Niigata}
\affiliation{Osaka City University, Osaka}
\affiliation{Osaka University, Osaka}
\affiliation{Panjab University, Chandigarh}
\affiliation{Peking University, Beijing}
\affiliation{Princeton University, Princeton NJ}
\affiliation{University of Science and Technology of China, Hefei}
\affiliation{Seoul National University, Seoul}
\affiliation{Sungkyunkwan University, Suwon}
\affiliation{University of Sydney, Sydney NSW}
\affiliation{Tata Institute of Fundamental Research, Bombay}
\affiliation{Toho University, Funabashi}
\affiliation{Tohoku Gakuin University, Tagajo}
\affiliation{Tohoku University, Sendai}
\affiliation{University of Tokyo, Tokyo}
\affiliation{Tokyo Institute of Technology, Tokyo}
\affiliation{Tokyo Metropolitan University, Tokyo}
\affiliation{Tokyo University of Agriculture and Technology, Tokyo}
\affiliation{Toyama National College of Maritime Technology, Toyama}
\affiliation{University of Tsukuba, Tsukuba}
\affiliation{Utkal University, Bhubaneswer}
\affiliation{Virginia Polytechnic Institute and State University, Blacksburg VA}
\affiliation{Yokkaichi University, Yokkaichi}
\affiliation{Yonsei University, Seoul}

\author{K.~Abe}               
\affiliation{High Energy Accelerator Research Organization (KEK), Tsukuba}
\author{K.~Abe}               
\affiliation{Tohoku Gakuin University, Tagajo}
\author{R.~Abe}               
\affiliation{Niigata University, Niigata}
\author{T.~Abe}               
\affiliation{Tohoku University, Sendai}
\author{I.~Adachi}            
\affiliation{High Energy Accelerator Research Organization (KEK), Tsukuba}
\author{Byoung~Sup~Ahn}       
\affiliation{Korea University, Seoul}
\author{H.~Aihara}            
\affiliation{University of Tokyo, Tokyo}
\author{M.~Akatsu}            
\affiliation{Nagoya University, Nagoya}
\author{Y.~Asano}             
\affiliation{University of Tsukuba, Tsukuba}
\author{T.~Aso}               
\affiliation{Toyama National College of Maritime Technology, Toyama}
% ...

\collaboration{Belle Collaboration}
\noaffiliation

\begin{abstract} 
\rm 
We have measured the branching fractions of the colour suppressed decays 
$B^{+} \rightarrow \psi(2S) K^{*+}(892)$ and $B^{0} \rightarrow 
\psi(2S)\,K^{*0}(892)$ using a data sample of 84 million $B\bar{B}$ events 
recorded by the Belle detector on the $\Upsilon(4S)$ resonance. The branching 
fractions for the charged and neutral mode decays are 
$(8.13 \pm 0.77 \pm 0.89) \times 10^{-4}$ and $(7.20 \pm 0.43 \pm 0.65) \times 10^{-4}$,  
respectively. 
\end{abstract} 
 \maketitle
\tighten
{\renewcommand{\thefootnote}{\fnsymbol{footnote}}}
\setcounter{footnote}{0}
\section{Introduction} 
The decays $B  \rightarrow \psi(2S)\,K^{*}$  are colour-suppressed two-body 
decays with two vector mesons in the final state.  Measurements of such 
processes can provide insight into the interplay between the weak and strong 
interactions. 

We study eight different decay channels, which are combinations of
two decay modes for the $\psi(2S)$: 
$\psi(2S) \rightarrow \ell \ell$ and $\psi(2S) \rightarrow  J/\psi\,\pi^{+}\pi^{-}$;    
and four different $K^*$ final states: $K^{*+} \rightarrow K^{+}\pi^{0}$, 
$K^{*+} \rightarrow K_{S}\pi^{+}$, $K^{*0} \rightarrow K^{-}\pi^{+}$, and
$K^{*0} \rightarrow K_{S}\pi^{0}$ (charged conjugate processes are implied throughout
this paper).  
The branching fractions for $B \rightarrow \psi(2S)\, K^{*}$  have been 
measured previously by the CLEO\cite{bfcleo} and CDF\cite{bfcdf} experiments. 
Our much larger data sample permits us to make more precise determinations 
of these quantities. 


\section{Data Sample and Event Selection} 
This measurement uses a data sample of approximately 84 million $B\bar{B}$ events,
corresponding to an integrated luminosity of 78~fb$^{-1}$, collected at the 
$\Upsilon$(4S) resonance by the Belle detector\cite{bldnim} at KEKB\cite{kekb}.  
KEKB is an energy-asymmetric double storage ring that collides 8~GeV electrons with 
3.5~GeV positrons.   


The Belle detector is a general-purpose large-solid-angle spectrometer with 
a 1.5-T solenoidal magnet.  It is designed to study the properties of 
$B$ mesons, charmed hadrons, and $\tau$ leptons, and also to study two-photon interactions. 
Charged particles are detected by a silicon vertex detector (SVD) that 
surrounds the interaction point, and then tracked by a cylindrical 
wire drift chamber (CDC). The CDC provides a three-dimensional momentum 
measurement of each track.  Charged particles are identified via information 
from silica aerogel Cherenkov counters (ACC), a time-of-flight 
scintillator barrel (TOF), and specific ionization ($dE/dx$) measurements 
in the CDC.  Electromagnetic showers are detected in an array of CsI(T$\ell$) 
crystals (ECL) that is located just inside the solenoidal coil.  Muons and 
$K_{L}$ mesons are identified by glass-electrode resistive-plate counters (KLM) 
embedded in the solenoid's magnetic flux return.

Hadronic events are selected from the raw data sample by requiring at least 
three good charged tracks emerging from near the $e^{+}e^{-}$ interaction point, 
within a cylindrical region of $r < 2.0$~cm 
and $|z| < 5.0$~cm, aligned along the positron beam axis, and at least 
one cluster of 0.1~GeV or more within the ECL barrel.  In the center-of-mass 
(CM) frame, the component of the total momentum along the beam axis of all 
charged tracks and neutral ECL clusters must be below $0.5\sqrt{s}/c$, where 
$\sqrt{s} = 10.58$~GeV is the total available energy. In the laboratory  
frame, the visible energy (the sum of the charged track momenta and the 
neutral cluster energies) must exceed $0.2\sqrt{s}$ and the neutral-only part 
must be between $0.1\sqrt{s}$ and $0.8\sqrt{s}$.  Continuum (non-resonant 
$e^{+}e^{-}\rightarrow q\bar{q}$)  events are suppressed relative to 
$e^{+}e^{-} \rightarrow \Upsilon(4S)$ by requiring that the ratio $R_{2}$ of 
the second to the zeroth Fox-Wolfram moments\cite{foxh2} falls below 0.5. 

\begin{figure}[h] 
\begin{center} 
\begin{tabular}{rl} 
   \epsfig{file=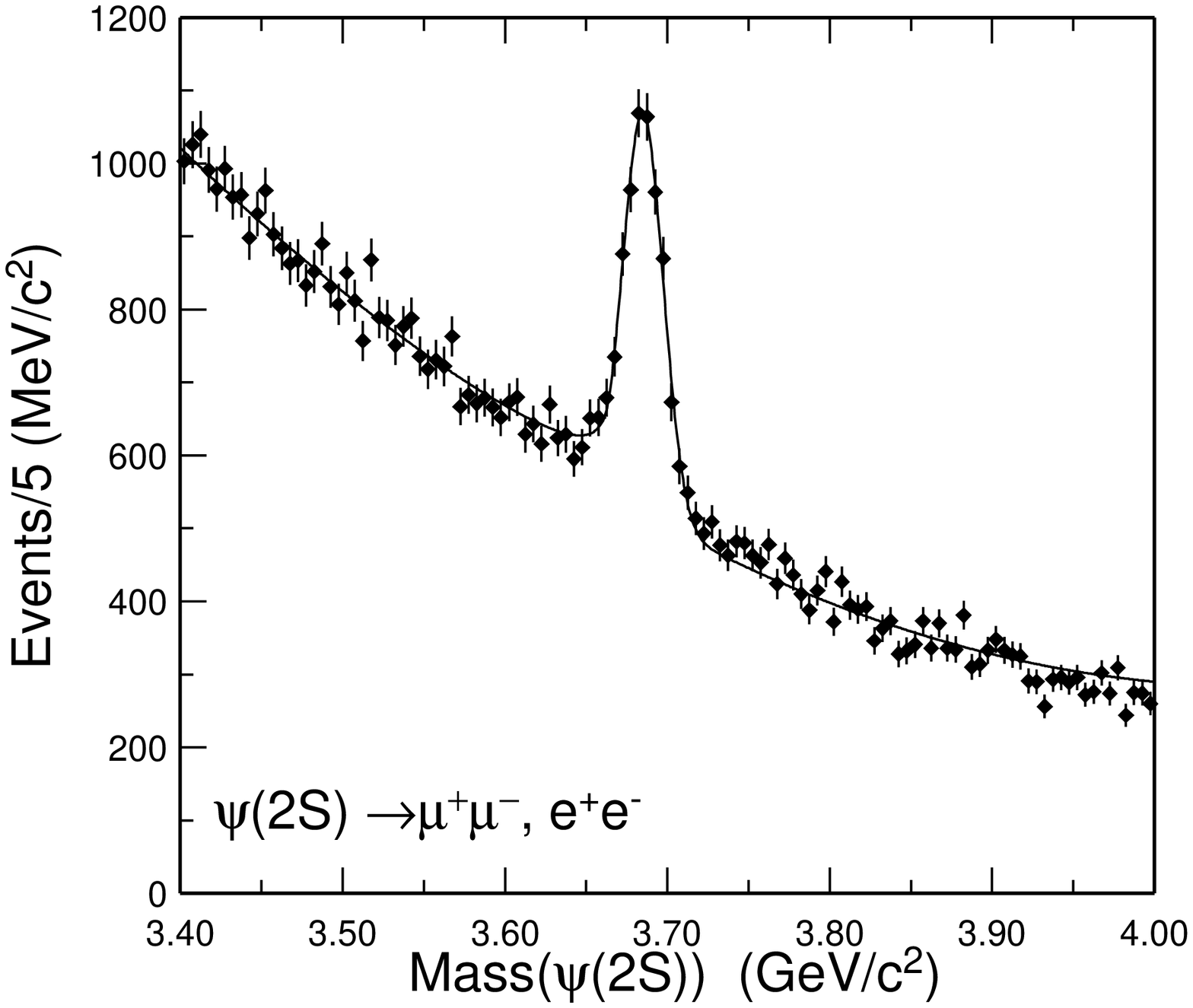,height=7.5cm}&
   \epsfig{file=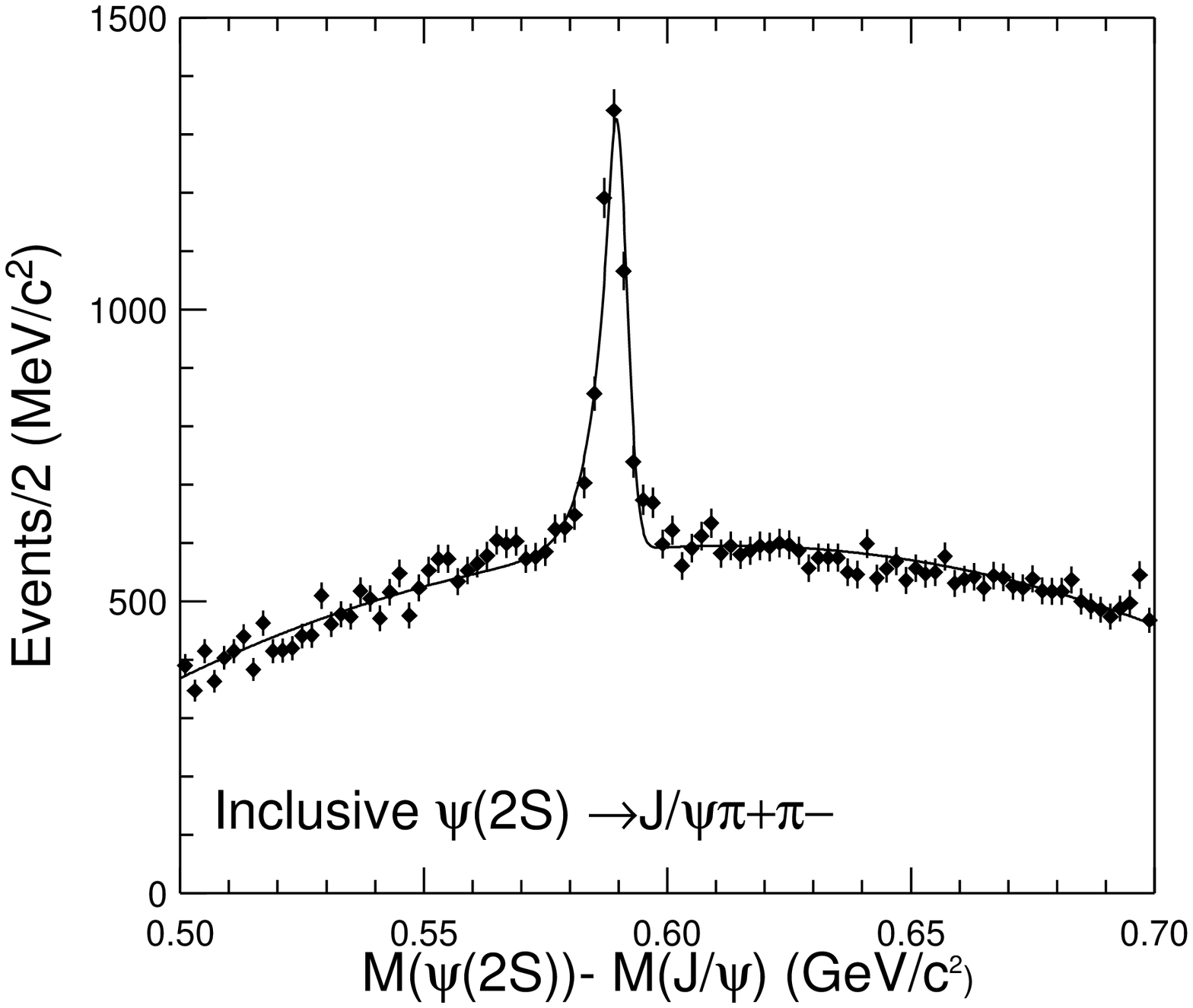,height=7.5cm}

\end{tabular}
\end{center} 
\caption{Invariant mass spectra of $\psi(2S)$ candidates.  The  
left plot shows $\psi(2S) \rightarrow \ell^{+}\ell^{-}$ while  right 
plot shows $\psi(2S) \rightarrow J/\psi\,\pi^{+}\pi^{-}$.
}
\label{figpsi2} 
\end{figure}

Candidate charmonium mesons are reconstructed through the decays 
$J/\psi \rightarrow \ell^+ \ell^-$, $\psi(2S) \rightarrow \ell^{+}\ell^{-}$, 
and   $\psi(2S) \rightarrow J/\psi \,\pi^{+}\pi^{-} \rightarrow 
\ell^{+} \ell^{-}\pi^{+}\pi^{-}$, where $\ell$ represents either an electron or a muon. 
Electrons are identified by a maximum-likelihood technique that uses energy and 
momentum information from the ECL and CDC, respectively.  If there is a neutral ECL cluster 
within $0.05$~rad of the electron/positron direction, it is assumed to 
arise from bremsstrahlung, and its energy is added to that of the lepton. 
Both tracks are required to originate from a common vertex by requiring that 
their distance of closest approach, measured along the beam axis, satisfies 
$|dz| < 5.0$~cm.   Candidate $J/\psi$ and $\psi(2S)$ mesons are 
retained if the mass difference between the reconstructed $e^{+}e^{-}$ invariant 
mass and the nominal $J/\psi$ or $\psi(2S)$ mass  falls in the range 
$ -0.150 < (M_{e^+e^-} - M_{J/\psi}) < 0.036$~GeV/$c^{2}$  or 
$ -0.186 < (M_{e^+e^-} - M_{\psi(2S)}) < 0.036$~GeV/$c^{2}$.
(The windows are wider on the low side of the charmonium mass peak to 
allow for residual bremsstrahlung.)
 Muons are identified by a maximum likelihood technique that uses the 
momentum information from the CDC in combination with the range and 
transverse scattering information from the KLM. 
As for electrons, muon pairs must satisfy the 
common-vertex requirement of $|dz| < 5.0\,{\rm cm}$. 
Candidate $J/\psi$ and $\psi(2S)$ mesons are retained if the reconstructed $\mu^+\mu^-$ 
mass satisfies 
$ -0.060 < (M_{\mu^+\mu^-} - M_{J/\psi}) < 0.036$~GeV/$c^{2}$  or 
$ -0.070 < (M_{\mu^+\mu^-} - M_{\psi(2S)}) < 0.036$~GeV/$c^{2}$.
The invariant-mass spectrum of the candidate inclusive $\psi(2S)$ sample is plotted in 
Fig.~\ref{figpsi2}.\\

Additional $\psi(2S)$ candidates are reconstructed through their decay 
to $J/\psi\,\pi^+\pi^-$.  The $J/\psi$ meson is reconstructed via its decay 
to dileptons, as described above.  Here, the momenta of the leptons and the 
parent $J/\psi$ candidate are redetermined in a mass-constrained fit 
assuming the nominal $J/\psi$ meson mass.  Pions are selected from charged 
tracks after electron and muon rejection.   To reduce combinatorial 
background, the $\pi^+\pi^-$ invariant mass must exceed 0.4~GeV/$c^2$, 
and the CM momentum of the $\psi(2S)$ candidate must fall below 
1.7~GeV/$c$.  The three-dimensional distance between the reconstructed 
vertices of the lepton pair and the pion pair must be less than 
$0.5\,{\rm cm}$.  We retain a candidate $\psi(2S)$ meson if its mass 
exceeds the reconstructed $J/\psi$ mass by $(0.59\pm 0.01)\,{\rm GeV}/c^2$. 
The spectrum of the mass difference between the $\psi(2S)$ 
candidate and the $J/\psi$ daughter is shown in Fig.~\ref{figpsi2}. 

We reconstruct $K^{*0}$ candidates from their decays to $K^{+}\pi^{-}$ or 
$K_{S}\pi^{0}$, and $K^{*+}$ candidates from their decays to $K^{+}\pi^{0}$ or 
$K_{S}\pi^{+}$.  Charged tracks are identified as kaons if they satisfy the 
kaon selection criterion derived from a maximum likelihood technique that 
distinguishes them from pions using information from the CDC, ACC, and the TOF; 
otherwise, they are classified as pions.  Neutral pions are reconstructed from pairs of 
photons (i.e., neutral ECL clusters) whose individual energies exceeds $60\,{\rm MeV}$ 
and whose invariant mass is in the window $0.12 < 
M_{\gamma\gamma} < 0.15$~GeV/$c^2$. 
(The photon and pion momenta of the surviving candidates are also recalculated 
assuming the nominal mass of the $\pi^0$ meson.) 

$K_S$ mesons are reconstructed from two oppositely charged pions whose 
distance of closest approach to the interaction point exceeds $0.25\,{\rm mm}$ 
in the plane perpendicular to the beam axis. 
If the two tracks have SVD hits, they must originate from a common vertex 
that is within the SVD but detached from the interaction point. 
Otherwise, the azimuthal coordinate of the common vertex must agree with 
the azimuthal direction of the pion pair within $0.1\,{\rm rad}$.  The 
invariant mass of the pion pair must be in the range 
$0.482\,{\rm GeV}/c^2 < M_{\pi\pi} < 0.515\,{\rm GeV}/c^2$.  

When combining a neutral pion with either a charged or neutral kaon to form a 
$K^{*}$ candidate, we require the  $K^{*}$ helicity angle 
(the angle of the kaon with respect to the $K^{*}$ direction in the $K^{*}$ 
rest frame) to satisfy $\cos\theta < 0.85$.   This requirement provides rejection of the 
slow-pion background.   The invariant mass of the $K^+\pi^-$ or $K_S\pi^+$ 
combinations and the $K_S\pi^0$ or $K^+\pi^0$ combinations must agree with the nominal 
$K^{*}$ meson mass within $0.075\,{\rm GeV}/c^2$ and $0.090\,{\rm GeV}/c^2$, 
respectively.   For the $K^{*0} \rightarrow K^+ \pi^-$ mode, the momenta of 
the kaon and pion are recalculated in a vertex-constrained fit. 

\section{Measurement of Branching Fractions} 
Candidate $B$ mesons are reconstructed by combining $\psi(2S)$ candidates 
with $K^{*}$ candidates.  For each $B$ candidate, we calculate the 
momentum $p_B$ and energy $E_B$ in the CM frame, and from these the beam-constrained 
mass $M_{bc} = \sqrt{ E_{\rm beam}^2 - p_B^2 }$ and the energy difference 
$\Delta E = E_B - E_{\rm beam}$.  ($E_{\rm beam} \equiv \sqrt{s}/2$ is the 
beam energy in the CM frame.)

If there are multiple $B$ candidates in an event, the best one is chosen 
by first discarding all but the best-reconstructed $\psi(2S)$ candidate, 
and then selecting the $B$ candidate with the smallest $\chi^{2}$. This quantity
is defined by $ \chi^2 = \sum_{i} (m_{i} - \mu_{i})^{2}/\sigma_{i}^{2}$, where  
$\mu_{i}$ is the central value of the measured parameter $m_{i}$---i.e., the mass of 
the reconstructed $\psi(2S)$ or $K^{*}$---and $\sigma_{i}$ the corresponding 
uncertainty.  The $B$ candidate selected in this way is retained only if it 
lies within a specified region of the $M_{bc}$--$\Delta E$ plane: 
$-0.2  < \Delta E <  0.2$~GeV and $5.2  < M_{bc} < 5.3$~GeV/$c^{2}$.  

A signal box, $|\Delta E| < 0.03$~GeV (0.04~GeV for $\pi^{0}$ modes) 
and $5.27 < M_{bc} < 5.29$~GeV/$c^{2}$, was determined using a large sample of 
Monte Carlo  signal events.   We have verified that the $M_{bc}$ and $\Delta E$ 
resolutions agree well between data and simulation.   In the cases where the 
final state contains only charged tracks ($\ell^+\ell^-K^+\pi^-$ and 
$\ell^+\ell^-\pi^+\pi^-K^+\pi^-$), the track momenta are recalculated assuming 
a common vertex constraint; this improves the resolutions in $M_{bc}$ and $\Delta E$ 
for signal $B^0$ candidates while having no effect on the background distribution.

\begin{figure} 
\begin{center} 
\begin{tabular}{rl} 
    \epsfig{file=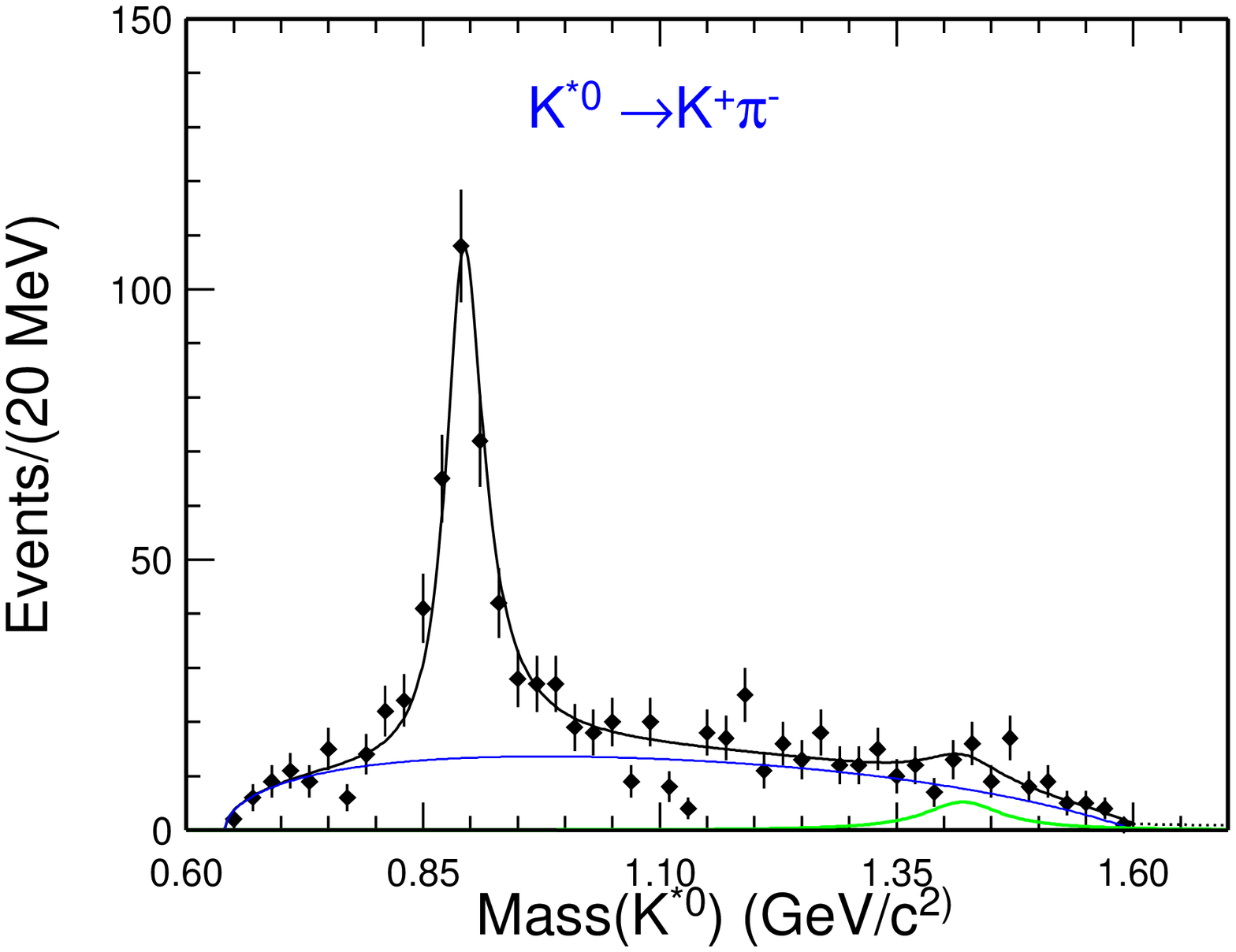,height=6cm}&
    \epsfig{file=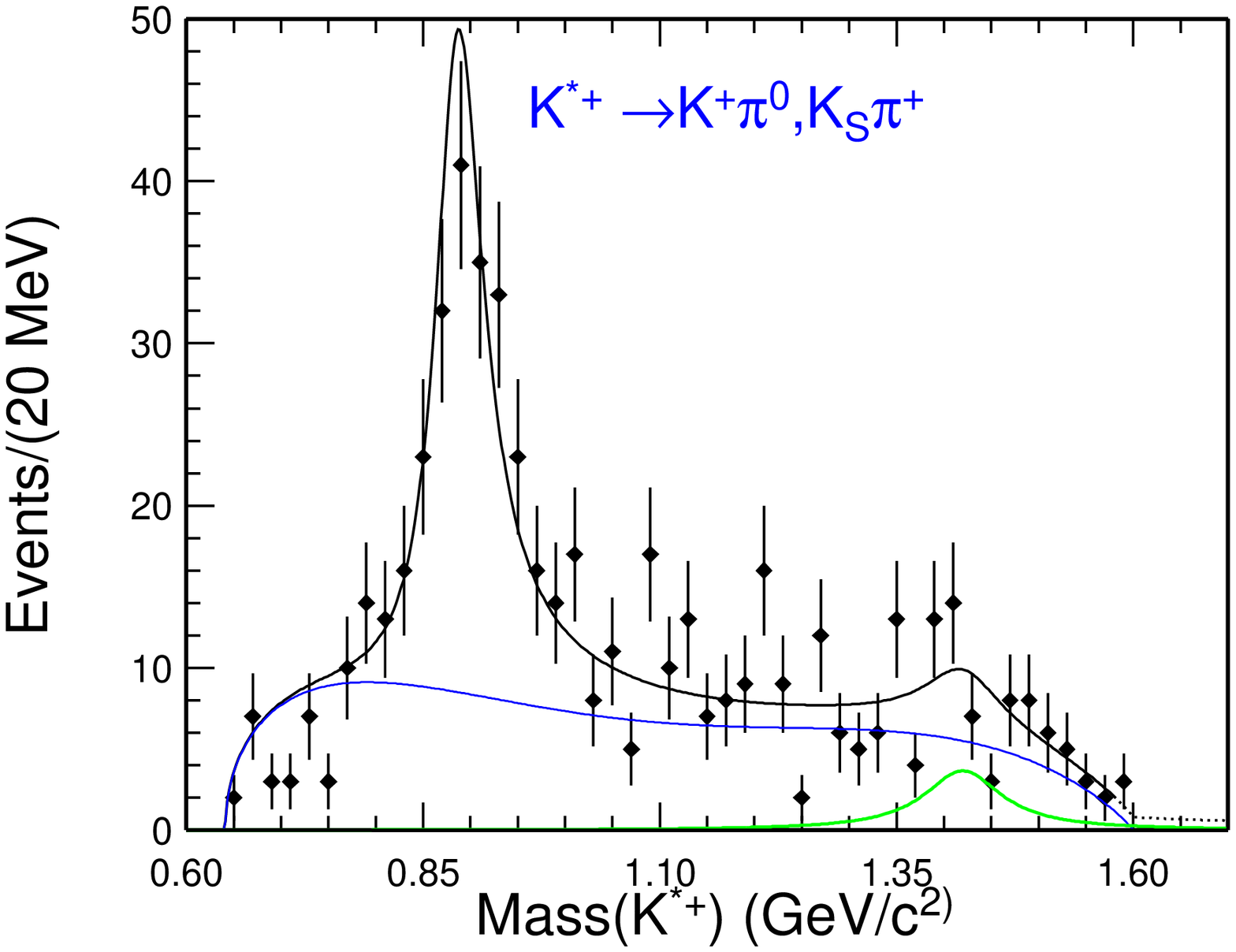,height=6cm}
\end{tabular}
\end{center} 
\caption{Invariant mass of the (a) $K^{*0}$ candidates reconstructed 
from $K^{-}\pi^{+}$ and  $K_{S}\pi^{0}$ pairs,  (b) $K^{*+}$ candidates 
reconstructed from $K^{+}\pi^{0}$ and $K_S\pi^{+}$ pairs.} 
\label{kstar} 
\end{figure}

Using the $M_{bc}$ and $\Delta E$  signal box cut to restrict the sample of $B$ 
meson candidates,  we obtain $K\pi$ mass distributions for $K^{*0}$ and $K^{*+}$ 
candidates.   These distributions are fitted  to Breit-Wigner $p$-wave functions 
for signal and second-order polynomial functions for background.  The $K\pi$ mass 
spectra of the exclusive $K^{*}$ candidates are shown in Fig.~\ref{kstar}.
$B \rightarrow \psi(2S) K^{*}$ backgrounds are classified as coming from  
the $q \overline{q}$ continuum, combinatorics, feed across, and 
non-resonant $K\pi$ decay.  The relative importance of each background 
category varies with the signal mode under consideration.  
Backgrounds of each type were studied for each mode using a combination 
of Monte Carlo samples and data sidebands, as described below.\\
\indent  
We simulate the continuum background using the proportion $1\,:\,1.6\,:\,1$ 
for $b\,:\,uds\,:\,c$ quarks.  A Monte Carlo sample with a size corresponding
to twice the data sample was generated and subjected to the selection criteria
outlined above.  Since no events survive our cuts,  we conclude that the continuum
contribution is negligible.\\  
Combinatorial backgrounds are analyzed by using the sidebands of 
the experimental data.  (A MC study was also considered, but owing
to complications arising from non-resonant $B \rightarrow \psi(2S) K\pi$ 
backgrounds and contributions from higher $K^*$ resonances, it was 
not adopted.)   We take the areas in the $\Delta E$-$M_{bc}$ plane given by 
$5.2 < M_{bc} < 5.27~\rm{GeV}/c^2$ and $ -0.2 <\Delta E < -0.05$~GeV 
and $ 0.05<\Delta E < 0.12$~GeV for our sideband analysis.  We exclude 
the $ 0.12<\Delta E < 0.2$~GeV  region from this study to avoid possible 
non-combinatorial contributions  from $\psi(2S) K$ and $\psi(2S) \pi$ decays.
\indent 
Background from feed across is studied by using exclusive 
signal MC data generated using the program EvtGen~\cite{EvtGen}.
The generated particles are tracked through the detector using
the program GEANT~\cite{GEANT}.  Eight 20,000-event samples, 
corresponding to the eight $ B \rightarrow \psi(2S) K^*$ 
modes studied, are used.   For each decay mode studied, feed across
backgrounds from the remaining seven modes are summed.   The total
amount of feed across relative to signal yields is 8.1-12.9\%, 
depending on the mode.\\
\indent
Finally,  backgrounds from  non-resonant $B \rightarrow \psi(2S) K\pi$ modes
are studied by using a sideband in the $K\pi$ mass spectrum with
masses  in the range
$1.1 < M_{K\pi} < 1.3$~~GeV/$c^2$.  We obtain an estimated background 
contribution from non-resonant $K\pi$ modes of about 
6.8\% for $\psi(2S)K^{*+}$ and 4.7\% for $\psi(2S)K^{*0}$.    
\begin{figure} 
\begin{center}
  \begin{tabular}{rl} 
    \epsfig{file=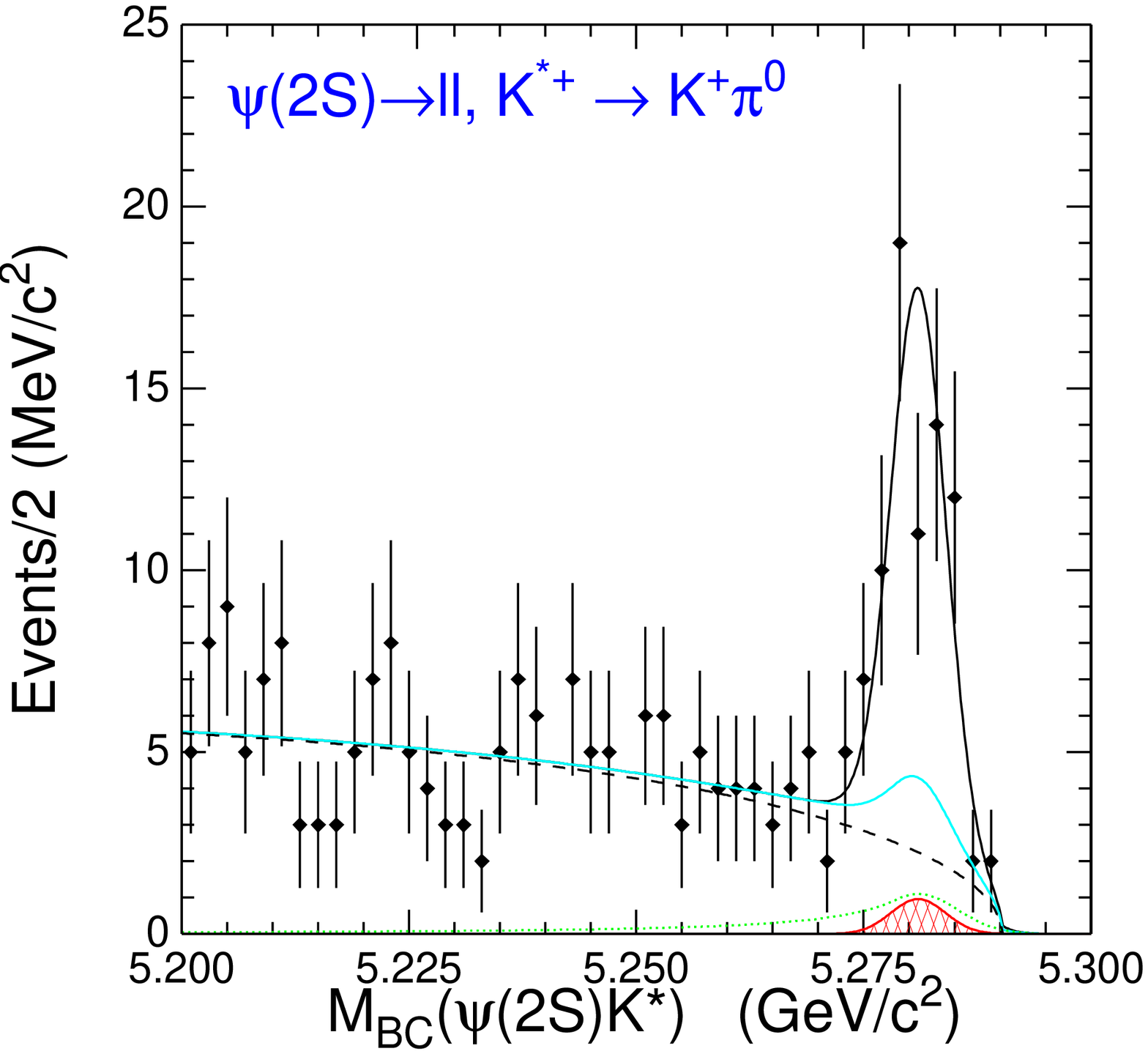,height=6.0cm,clip=}& 
    \epsfig{file=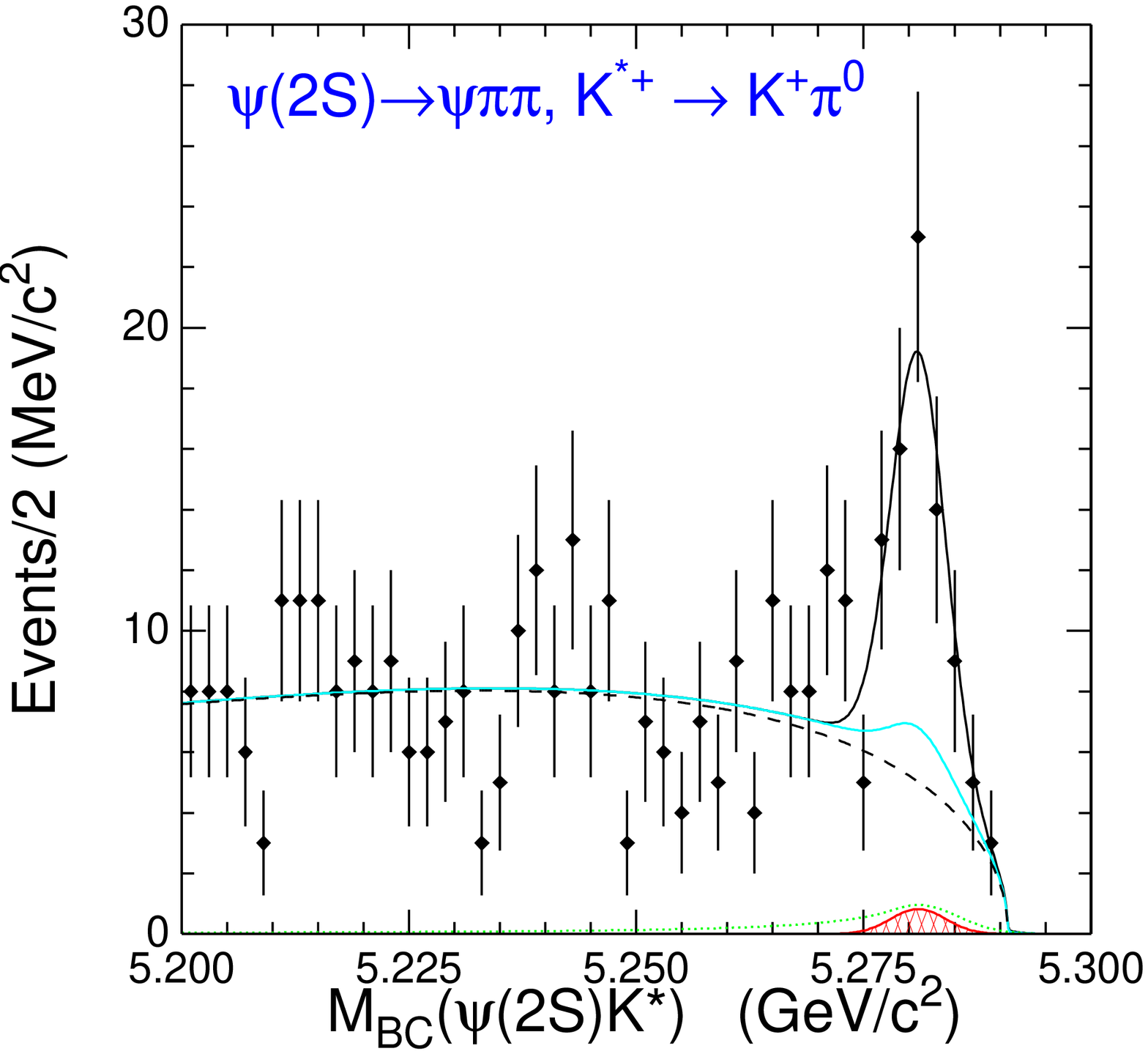,height=6.0cm,clip=} \\ 
    \epsfig{file=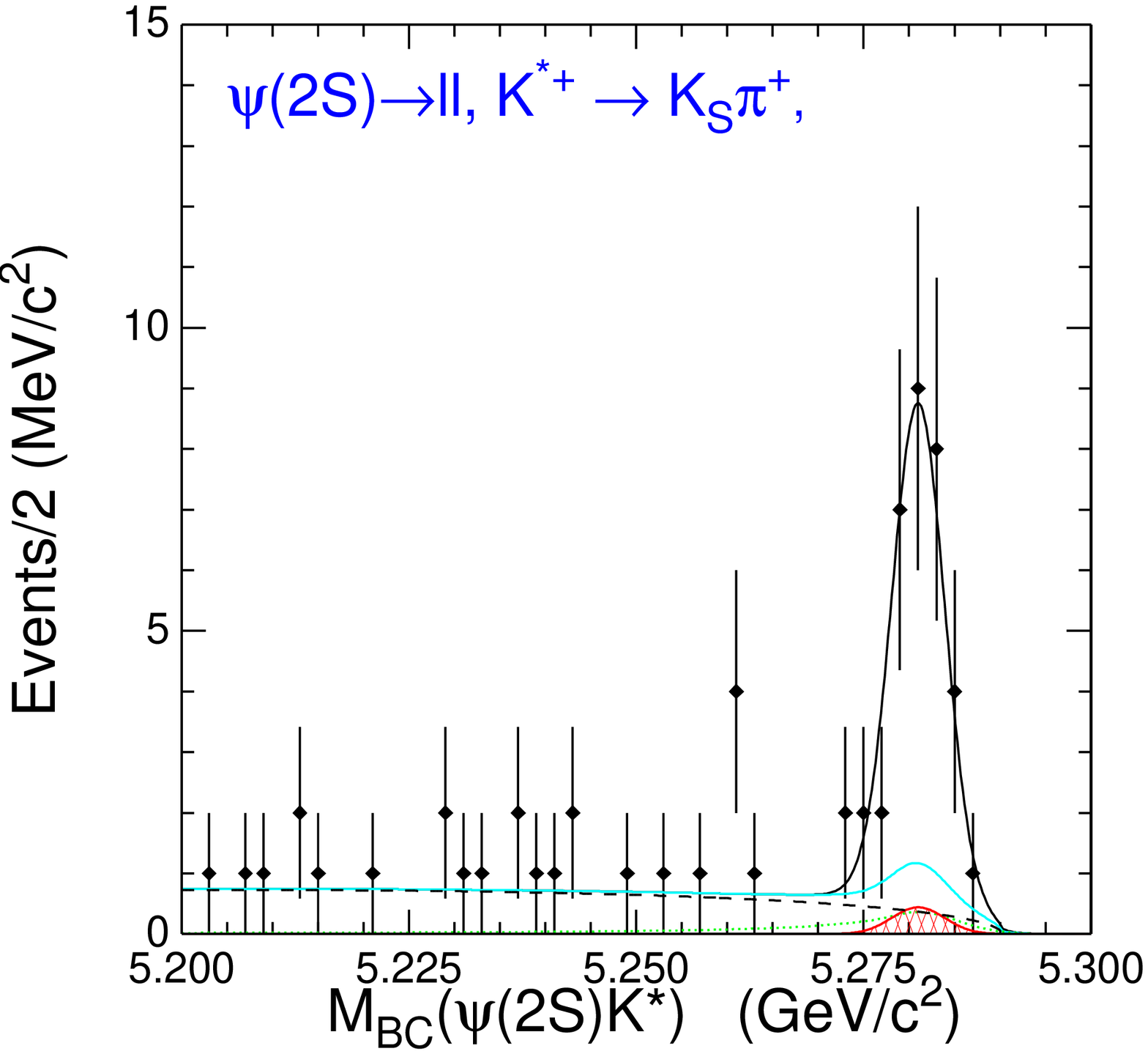,height=6.0cm,clip=}& 
    \epsfig{file=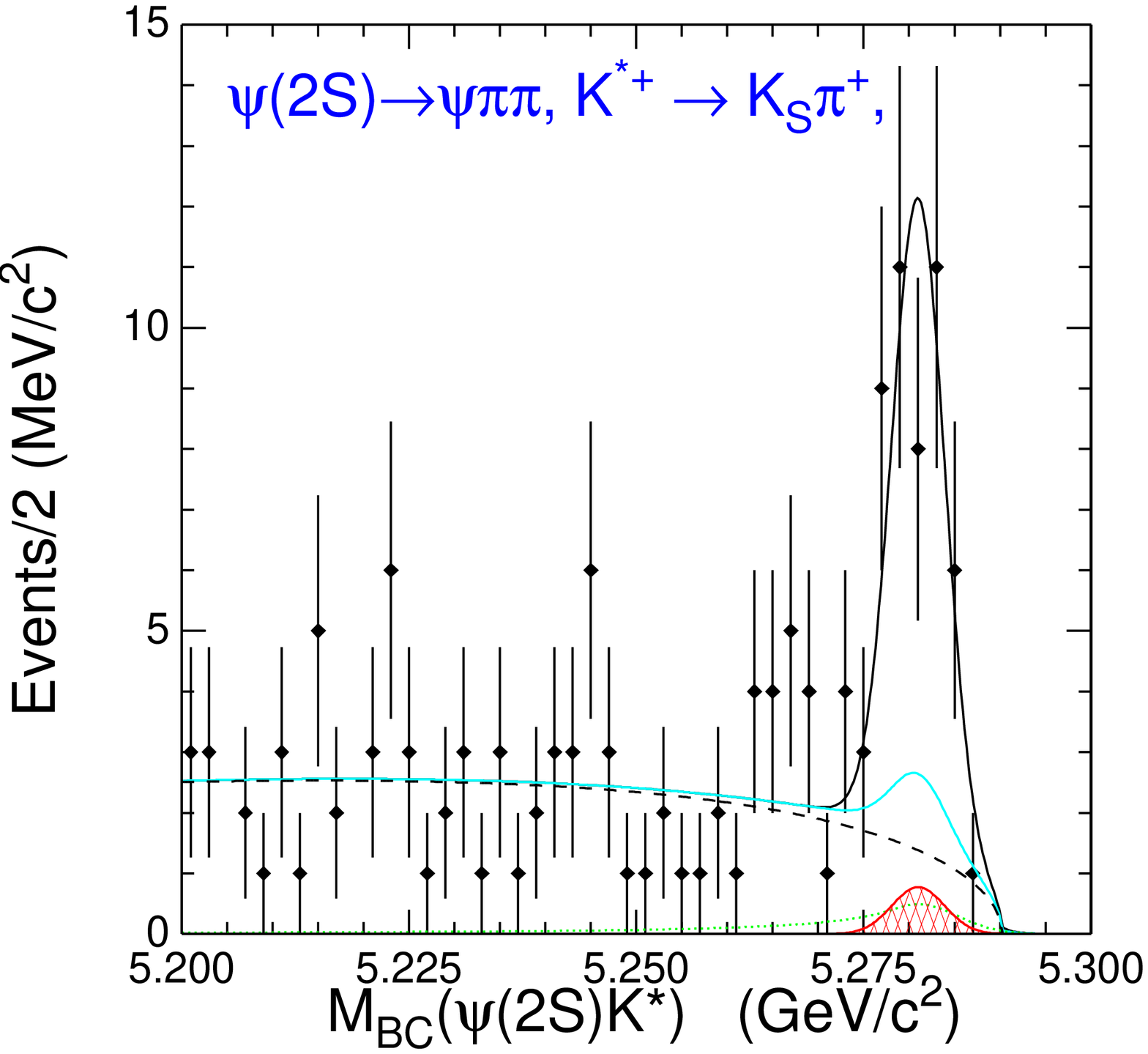,height=6.0cm,clip=}
  \end{tabular}  
\end{center} 
\caption{
Beam-constrained mass plots and fits for 
$B^{+} \rightarrow \psi(2S) K^{*+}$. The left-hand plots are for $\psi(2S) \rightarrow \ell^+ \ell^-$
and the right-hand plots are for $ \psi(2S) \rightarrow J/\psi \pi^+ \pi^-$.  The top plots 
show $K^{*+} \rightarrow K^{+}\pi^{0}$ and the bottom plots show 
$ K^{*+} \rightarrow K_{S}\pi^{+}$.  The dashed lines show the
combinatorial contribution to the background.  The dotted lines (green) show the 
feed across background and the solid lines with hatched areas (red) show the contributions from non-resonant decay. The gray lines give total background.} 
\label{mpsi2kst1}
\end{figure} 

\begin{figure}
\begin{center}
  \begin{tabular}{rl} 
    \epsfig{file=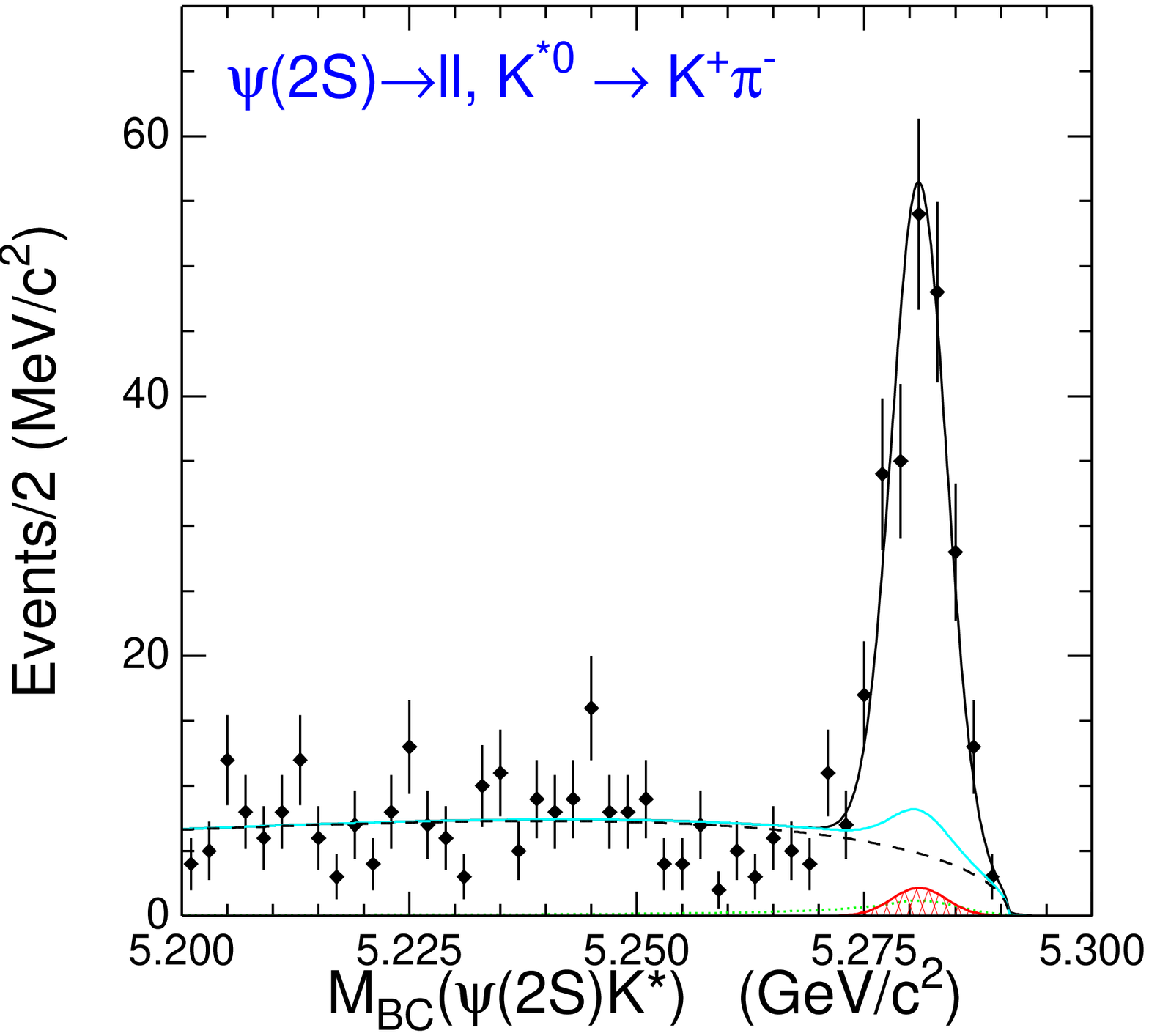,height=6.0cm,clip=}& 
    \epsfig{file=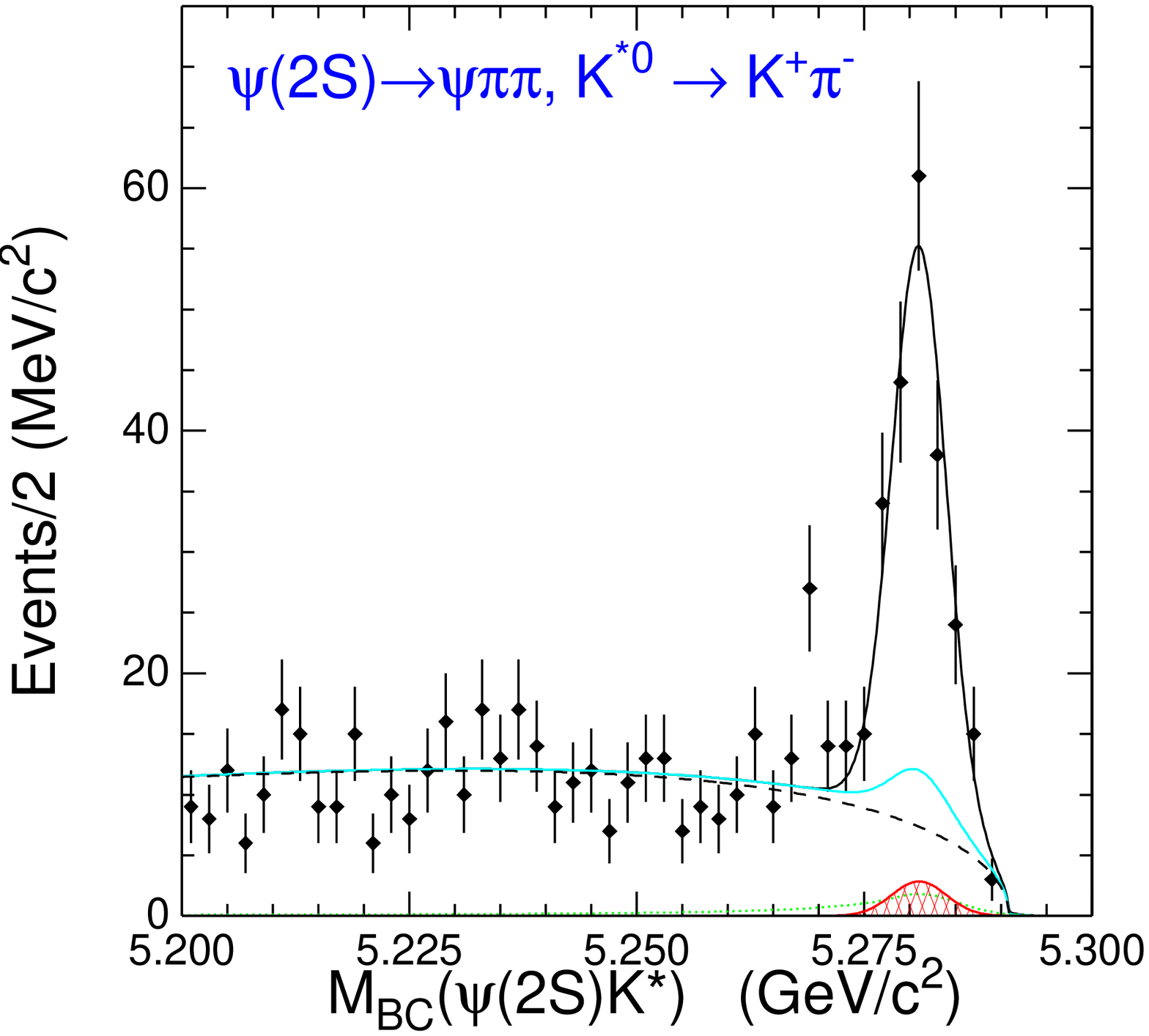,height=6.0cm,clip=}\\
    \epsfig{file=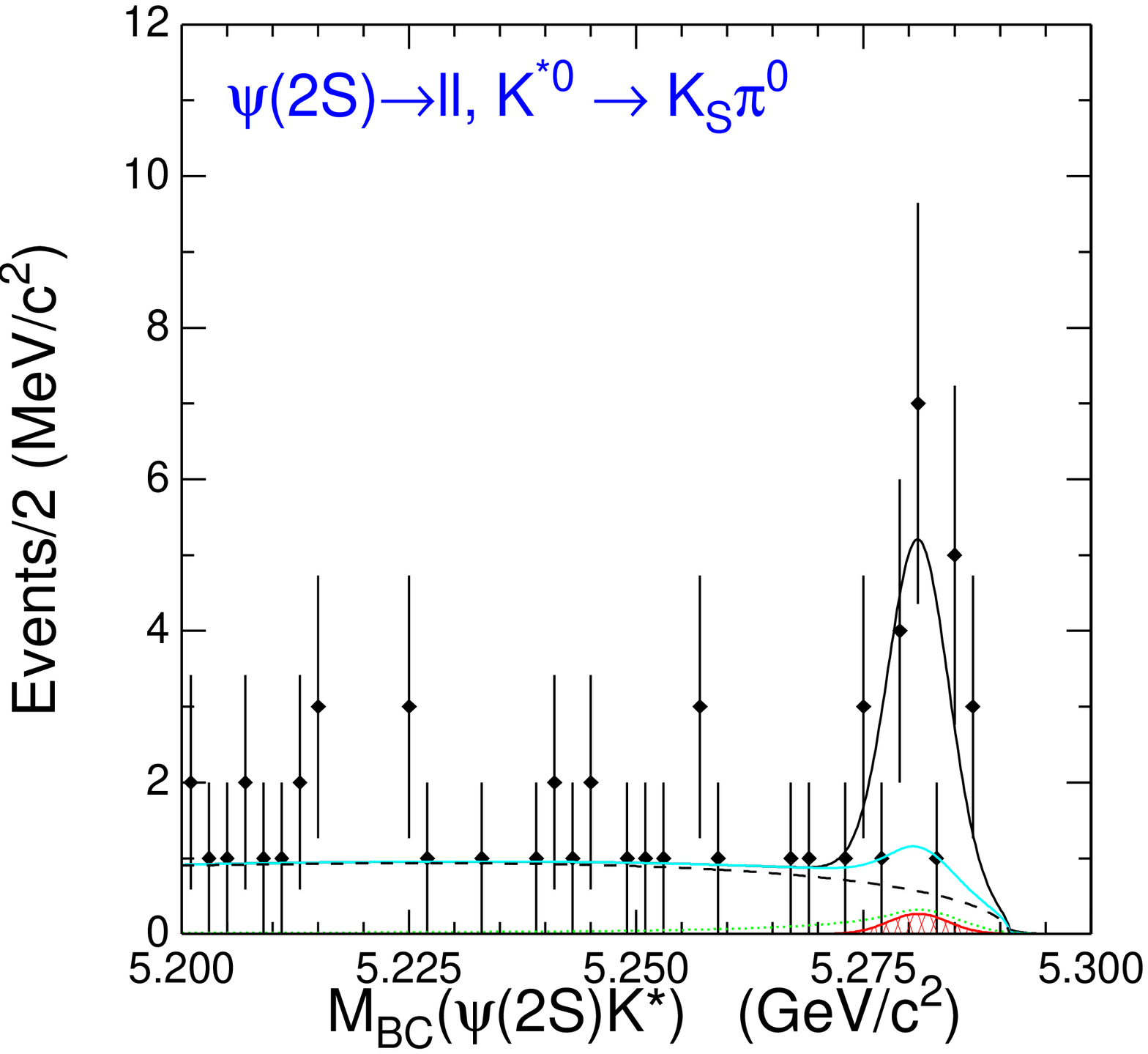,height=6.0cm,clip=}&
    \epsfig{file=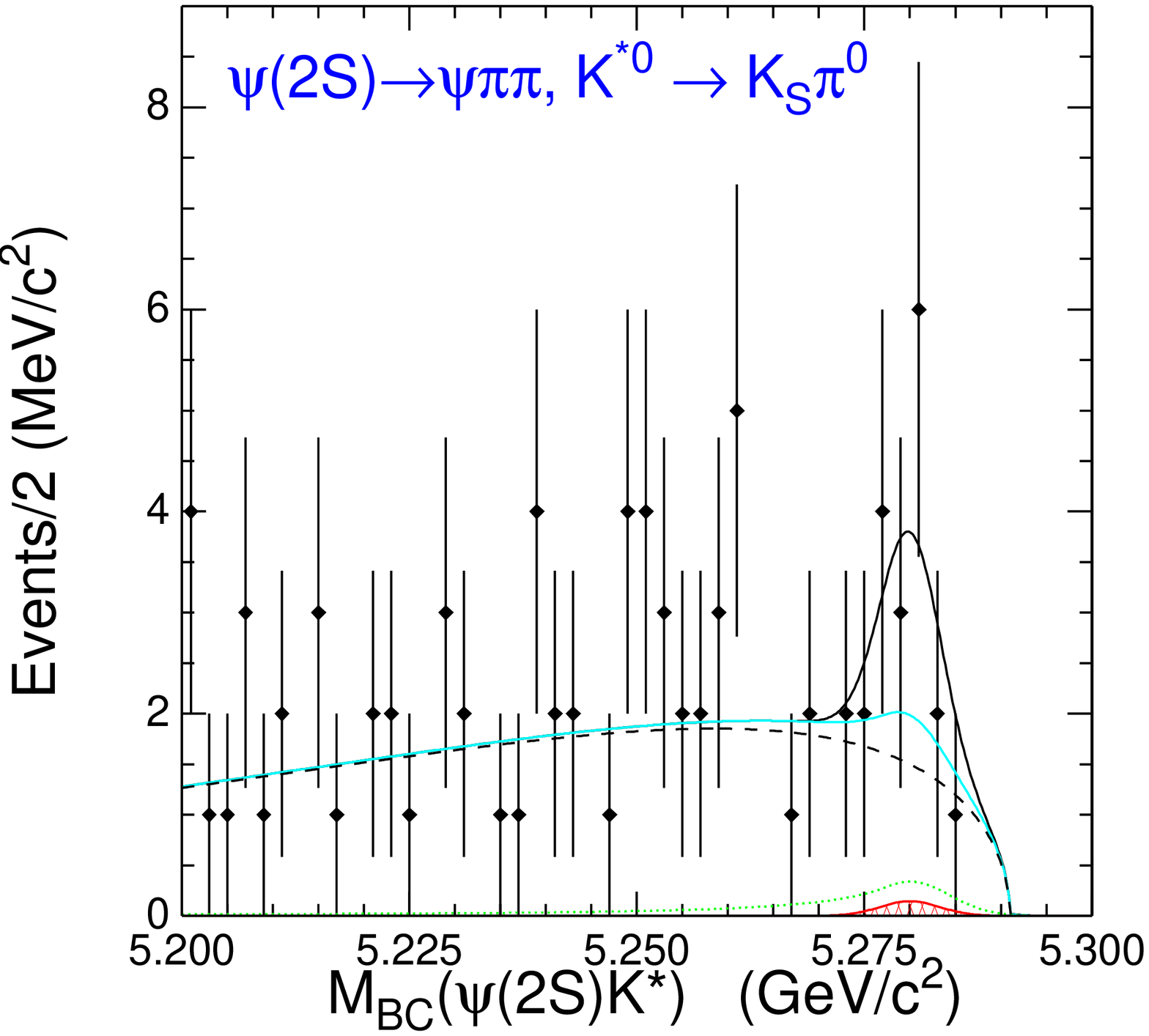,height=6.0cm,clip=}
  \end{tabular} 
\end{center} 
\caption{
Beam-constrained mass plots and fits for 
$B^0 \rightarrow \psi(2S) K^{*0}$. The left-hand plots are for $\psi(2S) \rightarrow \ell^+ \ell^-$
and the right-hand plots are for $ \psi(2S) \rightarrow J/\psi \pi^+ \pi^-$.  The top plots 
show $K^{*0} \rightarrow K^{+}\pi^-$ and the bottom plots show 
$ K^{*0} \rightarrow K_{S}\pi^0$.  The dashed lines show the
combinatorial contribution to the background.  The dotted lines (green) show the 
feed across background and the solid lines with hatched areas (red) show the contributions from non-resonant decay. The gray lines give total background.} 
\label{mpsi2kst2} 
\end{figure}

Measurements of the branching fractions are based on fits to the 
beam-constrained mass ($M_{bc}$) distributions.   Separate fits are
performed for each of the eight submodes under study.  These distributions
comprise signal and background events as discussed in the previous sections.
In the fits, the signal is described by a Gaussian whose mean and width
are determined from MC.  The combinatorial background is described using 
an ARGUS function~\cite{ARGUS} whose shape parameters are fixed using sideband data.
The feed across background is modeled by a Crystal Ball function~\cite{CB}
where both the shape and the level are fixed by MC.  The level of the feed across
background depends on the $B \rightarrow \psi(2S)K^*$ branching fractions
being measured, so an iterative procedure is adopted wherein the branching
ratios obtained using initial estimates for the feed across modes are
used to carry out a second analysis.   Since the feed across backgrounds are
a small fraction of the signal, this procedure quickly converges.  
The non-resonant
$K\pi$ background is described by a Gaussian of fixed shape and amplitude.
The level of the combinatorial background is allowed to vary.  
Figs.~\ref{mpsi2kst1} and \ref{mpsi2kst2}   
show the fit results for $B^{+} \rightarrow 
\psi(2S) K^{*+}$ and $B^0 \rightarrow \psi(2S) K^{*0}$ decays, respectively. 
Fitting results from the $M_{bc}$ distributions are checked by carrying out
a second set of fits to the $\Delta E$ distributions.  This method yielded
consistent results, with yield differences varying from 1 to 3\%. 
The reconstruction efficiency for each decay mode is  obtained using signal MC 
data  with the same cut criteria.     Table~\ref{tneb} shows the number of signal events,  
the reconstruction efficiency and the total number of background events for 
each decay mode. World average values~\cite{PDG} are used for the branching fractions for 
$ \psi(2S) \rightarrow \ell^+\ell^-$,  $ \psi(2S) \rightarrow J/\psi\,\pi^{+}\pi^{-}$,
$J/\psi \rightarrow \ell^+ \ell^-$, and $K^* \rightarrow K\pi$ decays. 
\begin{table} 
\begin{center} 
\begin{tabular}{p{3.0cm} c p{2.2cm} p{1.5cm} l l p{3.cm}}\hline\hline
Decay channel~& ~~$\psi(2S)$~~ & ~~$N_{sig}$ &$\varepsilon_{rec}$(\%)~~&$N_{bkg}$~& ~~~~${\cal B}\times 10^{4}$\\\hline 

$\psi(2S)K^{*+}(K^{+}\pi^0)$& $\ell\ell$  &  $  50.6 \pm 8.6 $&  15.85 & 33.2                      & $ 7.96 \pm 1.41 \pm 0.84 $~~\\
                      \vfill& $\psi\pi\pi$&  $  47.6 \pm 9.8 $&   5.25 & 57.1                  ~~~ & $ 8.97 \pm 1.78 \pm 0.96 $\\
$\psi(2S)K^{*+}(K_{S}\pi^{+})$& $\ell\ell$& $  32.2 \pm 6.0  $&  15.5  &  9.8                       & $ 7.69 \pm 1.46 \pm 0.82 $~~\\
                      \vfill & $\psi\pi\pi$ &  $  36.1 \pm 7.0 $&   6.42 & 18.2                 ~~ & $ 8.21 \pm 1.66 \pm 1.12 $\\
$\psi(2S)K^{*0}(K^{+}\pi^{-})$ & $\ell\ell$& $ 181.4 \pm 15.2 $&  29.5  & 62.6                      & $ 7.66 \pm 0.62 \pm 0.61 $~~ \\
                      \vfill  & $\psi\pi\pi$&  $ 164.2 \pm 14.6 $&  12.1  & 91.6                ~~~ & $ 6.70 \pm 0.65 \pm 0.74 $\\
$\psi(2S)K^{*0}(K_{S}\pi^{0})$& $\ell\ell$& $  13.2 \pm 4.2  $&  10.3  &  7.3                   ~~~~& $ 9.35 \pm 2.73 \pm 0.97 $~~~ \\
                      \vfill & $\psi\pi\pi$ &  $   7.9 \pm 4.3 $&   4.01 & 13.4                ~~~ & $ 5.75 \pm 3.43 \pm 1.02 $ \\
\hline\hline
\end{tabular} 
\end{center} 
\caption{Number of signal events, reconstruction efficiency, number of 
background events and measured branching fractions  for each decay mode.} 
\label{tneb} 
\end{table}  
Systematic errors  arise from uncertainties in detector efficiency, particle 
identification efficiency,  reconstruction efficiency, and beam-constrained mass 
fitting.  We estimate these as follows: tracking efficiency (1.0\% per track);
particle identification efficiency (3.0\% for leptons and less than 1\% for kaons 
and pions);  background contribution (5.0 - 7.0\% decay mode dependent);  neutral 
pion reconstruction (4.0\%);   number of $B\bar{B}$ events (1.0\%);   
branching fraction for secondary decay (1.7\%); and the effect of polarization   (1.7\%).
The uncertainty in the background, which is the dominant contribution
to the overall systematic error, 
is due to uncertainties surrounding the contributions of higher-mass
$K^*$ resonances to the $1.1 < M_{K\pi}< 1.3$~GeV/$c^2$ mass region used to 
estimate the non-resonant contributions to the $K\pi$ mass region under
the $K^*(890)$ peak.  

The measured branching fractions are summarized in  Table \ref{tneb}.  
Table~\ref{brft2} shows the branching ratios that result from the weighted average of the
four final states for the charged and neutral mode decays.  Table~\ref{brft2}
also shows previous measurements by CLEO~\cite{bfcleo} for comparison.  
\begin{table}[h] 
\begin{center} 
\begin{tabular}{p{3.5cm} p{3.5cm} p{3.5cm}}\hline \hline 
\vfill
Decay channel & Belle (${\cal{B}}\times 10^{4})$& CLEO (${\cal{B}}\times 10^{4})$\\\hline 

  $B^{+}\rightarrow\psi(2S)K^{*+} $ & $ 8.13 \pm 0.77 \pm 0.89 $ & $ 8.9 \pm 2.4 \pm 1.2 $ \\ 
  $B^{0}\rightarrow \psi(2S)K^{*0}$ & $ 7.20 \pm 0.43 \pm 0.65 $ & $ 7.6 \pm 1.1 \pm 1.0 $\\ 

\hline \hline 
 \end{tabular}
\end{center} 
\caption{The results of measured branching fractions after taking weighted average for different decay channels.} 
\label{brft2} 
\end{table}
\section{conclusion}
The branching fraction for $B\rightarrow\psi(2S)K^{*}(890)$  has been measured using a sample
of 84 million $B$ pairs.  The results are consistent with earlier measurements from CLEO~\cite{bfcleo}
and CDF~\cite{bfcdf}, but offer improved statistical accuracy.  
 

\begin{thebibliography}{20} 
\bibitem{bfcleo} S.J. Richichi, {\it et al.}(CLEO Collaboration)  Phys. Rev. {\bf D63 }
(2001) 031103 
\bibitem{bfcdf}  F. Abe, {\it et al.} (CDF Collaboration) Phys.Rev.{\bf D58} (1998)  072001 
\bibitem{bldnim} A.Abashian {\it et al.}, Belle Collaboration Nucl. Instr. and Meth. A  {\bf 476} (2002) 117. 
\bibitem{kekb} E. Kikutani ed., KEK Preprint 2001-157 (2001), to be published in NIM.
\bibitem{foxh2} G. C. Fox and S.Wolfram, Phys. Rev. Lett. {\bf 41} (1978) 1581 
(2000) 2886. 
\bibitem{EvtGen} D.J.~Lange, Nucl. Instr. and Meth. {\bf A462} (2001) 152.   See also
EvtGen homepage {\tt http://www.slac.stanford.edu/$\sim$lange/EvtGen/ }.
\bibitem{GEANT} R.~Brun {\it et al.}, GEANT 3 Manual, CERN Program Library Long Writeup W5103, 1994.
\bibitem{ARGUS} ARGUS Collaboration, H.~Albrecht {\it  et al.}, Phys. Lett. {\bf B241} (1990) 278.
\bibitem{CB} M.~Oreglia, Ph.D. Thesis, Stanford University, Report No. SLAC-236 (1980).  
\bibitem{PDG} Particle Data Group, K.~Hagiwara {\it et al.}, Phys. Rev. {\bf D66} (2002) 010001-1.

\end{thebibliography}
\end{document}